\documentstyle[10pt,twocolumn,amssymb,graphicx,amsmath,graphicx,psfig,caption,subcaption,cite,float,setspace,xcolor]{IEEEtran}

\newcommand{\bi}{\begin{itemize}}
\newcommand{\ei}{\end{itemize}}

\newcommand{\be}{\begin{enumerate}}
\newcommand{\ee}{\end{enumerate}}
\newcommand{\bd}{\begin{description}}
\newcommand{\ed}{\end{description}}
\newcommand{\bc}{\begin{center}}
\newcommand{\ec}{\end{center}}
\newcommand{\bt}{\begin{tabbing}}
\newcommand{\et}{\end{tabbing}}
\newcommand{\bfig}{\begin{figure}}
\newcommand{\efig}{\end{figure}}
\newcommand{\beq}{\begin{equation}}
\newcommand{\beqarr}{\begin{eqnarray}}
\newcommand{\beqarrn}{\begin{eqnarray*}}
\newcommand{\eeq}{\end{equation}}
\newcommand{\eeqarr}{\end{eqnarray}}
\newcommand{\eeqarrn}{\end{eqnarray*}}
\newcommand{\bflr}{\begin{flushright}\vspace{-0.2in}}
\newcommand{\eflr}{\end{flushright}}
\newcommand{\bsub}{\begin{subequations}}
\newcommand{\esub}{\end{subequations}}
\newcommand{\barr}{\begin{array}}
\newcommand{\earr}{\end{array}}
\newcommand{\nn}{\nonumber}

\def\binom#1#2{\left( \! \! \barr{c} #1 \\ #2 \earr \! \! \right)}

\def\undb#1{\mbox{\bf{#1}}}

\def\dn{\stackrel{\scriptscriptstyle \triangle}{=}}

\def\arg{\mbox{arg}}

\begin{document}
\title{RIS-Assisted Space Shift Keying with Non-Ideal Transceivers and Greedy Detection}
\author{Aritra Basu, Soumya~P.~Dash,~\IEEEmembership{Senior Member,~IEEE}, and~Sonia A\"{i}ssa,~\IEEEmembership{Fellow,~IEEE}
\thanks{A. Basu and S. A\"{i}ssa are with the  Institut National de la Recherche Scientifique (INRS), Montreal, QC H5A 1K6, Canada; e-mail: \{aritra.basu, sonia.aissa\}@inrs.ca.}
\thanks{S. P. Dash is with the School of Electrical and Computer Sciences, Indian Institute of Technology Bhubaneswar, Argul, Khordha, 752050 India; e-mail: soumyapdashiitbbs@gmail.com.}
\thanks{This work was supported in part by a Discovery Grant from the Natural Sciences and Engineering Research Council (NSERC) of Canada.}
}
\maketitle
\begin{abstract}
Reconfigurable intelligent surfaces (RIS) and index modulation (IM) represent key technologies for enabling reliable wireless communication with high energy efficiency. However, to fully take advantage of these technologies in practical deployments, comprehending the impact of the non-ideal nature of the underlying transceivers is paramount. In this context, this paper introduces two RIS-assisted IM communication models, in which the RIS is part of the transmitter and space-shift keying (SSK) is employed for IM, and assesses their performance in the presence of hardware impairments. In the first model, the RIS acts as a passive reflector only, reflecting the oncoming SSK modulated signal intelligently towards the desired receive diversity branch/antenna. The second model employs RIS as a transmitter, employing $M$-ary phase-shift keying for reflection phase modulation (RPM), and as a reflector for the incoming SSK modulated signal. Considering transmissions subjected to Nakagami-$m$ fading, and a greedy detection rule at the receiver, the performance of both the system configurations is evaluated. Specifically, the pairwise probability of erroneous index detection and the probability of erroneous index detection are adopted as performance metrics, and their closed-form expressions are derived for the RIS-assisted SSK and RIS-assisted SSK-RPM system models. Monte-Carlo simulation studies are carried out to verify the analytical framework, and numerical results are presented to study the dependency of the error performance on the system parameters. The findings highlight the effect of hardware impairment on the performance of the communication system under study. More specifically, minimal to no impact is observed on the system performance for low average SNR, while the effect is prominent for higher average SNR. The study emphasizes using such greedy detectors, which are robust to such practical system impairments.
\end{abstract}
\section{Introduction}
The fifth-generation (5G) of wireless communication has brought about a fresh perspective on mobile communications, broadly addressing three use cases: enhanced mobile broadband, ultra-reliable and low-latency communications, and massive machine-type communications \cite{9706187}. However, meeting the diverse requirements of these use cases with a single technology has proven to be quite challenging. This has inspired researchers to explore new techniques to prepare communication systems for such practical and futuristic applications. While the sixth-generation (6G) appears to build upon its 5G predecessor, the introduction of new user requirements, applications, use cases, and networking trends, are expected to present new challenges that necessitate the development of new communication paradigms, particularly at the physical layer \cite{8766143, 8869705}. To address this, unconventional paradigms are gaining interest, aiming to intelligently control the properties of the propagation medium, such as scattering, reflection, and refraction. Recently, this has been supported by the swift progress in radio frequency micro-electro-mechanical systems that has led to the development of reconfigurable intelligent surfaces (RISs) consisting of programmable and reconfigurable meta-surfaces \cite{fdh20, dixcy29}.

The distinctive characteristic of RISs lies in their ability to modify the propagation medium between the transmitter(s) and the receiver(s). The unique nature of RISs, coupled with their ease of deployment in real-world scenarios, render them highly practical and suitable to integrate into future communication systems \cite{lopp234, res29, pops22, MaBhPa:14}. The uniqueness of RISs also resides in their passive nature, which aids in achieving channel hardening similar to traditional multiple-input multiple-output (MIMO) techniques. Additionally, RIS-assisted systems exhibit lower implementation costs and complexity compared to MIMO communication systems  \cite{falturef2, dixcy}. Moreover, the reflective properties of these surfaces can be manipulated through software, leading to their conceptualization as software-defined surfaces \cite{dswqrsd}.

Some studies have also highlighted the complementary use of RISs and relays to exploit the advantages of both in beyond-5G network architectures \cite{cvgqyv}. The use of multiple or cascaded RISs to extend coverage and achieve quality-of-service was highlighted in \cite{dixcy29, london}. Furthermore, the authors in \cite{mkij} suggested enhancements to RIS-based systems through the implementation of several modulation schemes in symbiotic communications. The authors in \cite{dvsa} delved into the examination of the ergodic capacity of a RIS-assisted multi-user multiple-input single-output communication system, considering practical statistical channel state information (CSI). Moreover, the paper \cite{dsbaa} demonstrated the integration of RIS in terahertz communication by applying a deep reinforcement learning method. More recently, researchers explored using RIS as a part of the transmitter and showed significant merits as highlighted in \cite{kolkata, sbz, 10018169}.

Considering the merits of index modulation (IM) techniques, including the favorable trade-offs between spectral efficiency and energy efficiency that the systems can achieve, researchers have recently explored different RIS-aided IM communication models. IM refers to a family of modulation techniques that incorporates activation states to embed information in various domains including space, time, frequency, or a combination thereof \cite{turkiye29}. The fundamental principle of IM consists in the partitioning of the information bits into index bits and constellation bits, with the index bits ascertaining the active radio resources (antennas and sub-carriers), while the constellation bits are utilized for mapping the constellation symbols to be carried by the active radio resources.

Within this category, spatial modulation (SM), and space-shift keying (SSK) as a special case of SM, leverage the spatial-constellation diagram for the data modulation, providing low-complexity solutions for MIMO systems that surpass conventional modulation schemes \cite{dixcydoctor}. While SM and SSK both operate on the same fundamental idea, SM-based systems are more complex due to the additional requirement of an amplitude/phase modulation technique. In SSK, the transmitter structure is simpler due to the non-requirement of additional modulation schemes, making it suitable for meeting the low-power and low-complexity requirements of IoT applications \cite{dfdgndn}. Furthermore, the study in \cite{vetu2910} demonstrates an improved system efficiency and performance of an RIS-assisted SSK system as compared to an SSK system without using RIS. The RIS intelligently reflects the incoming SSK modulated signal, by optimizing the phase introduced by the RIS with the help of real-time CSI, thereby improving the quality of the received signal.

Recently, a RIS-assisted SSK communication model has been proposed in \cite{sbz}, where the RIS acts as a transmitter in addition to reflecting the incoming signal from the source. This design reduces the information load on the transmitter compared to generic SM modulation systems by eliminating the need for complex oscillators and phase shifters. The sensor-embedded RIS controller collects environmental data, such as temperature and humidity, which is then modulated using discrete $M$-ary phase-shift keying (PSK) symbols. These symbols are appended to the impinging SSK-modulated signals and transmitted as discrete phase shifts, known as reflection phase modulation (RPM), thus resulting in a RIS-assisted SSK-RPM scheme.

Though RIS-assisted SSK and RIS-assisted SSK-RPM communication systems hold a great potential \cite{MaBhPa:14, kolkata}, their true capabilities can only be understood by analyzing their performance in real-world conditions, particularly when the system transceivers are non-ideal \cite{henry}, e.g., due to distortion noise \cite{dfas}.

Motivated by this, we consider a RIS-aided receive diversity IM system, where the transceiver pair is affected by practical hardware impairments. Two system models are accounted for in this study. In the first model, the transmitter uses SSK modulation to encode data with the RIS functioning only as a passive reflector. For the second model, the RIS acts as a transmitter for $M$-PSK modulated symbols in addition to functioning as a reflector for the impinging SSK-modulated signal from the transmitter. The performance of these systems is evaluated for the symbols being transmitted via Nakagami-$m$ fading channels and the receiver employing a greedy rule for detecting the index of the receive diversity branch with the highest instantaneous received energy. Using the probability of erroneous index detection (PED) as a reliability metric, where we consider the erroneous selection of a receive diversity branch at the receiver over the other receive diversity branches, by utilizing the optimum RIS phase shifts for the considered receive diversity branch.
The ensuing contributions consist in:
\begin{itemize}
     \item Finding the exact closed-form expressions for the PED reliability metric, by using a characteristic function (CF) based approach.
    \item Deriving closed-form expressions for the bit error rate (BER) in both communication systems, based on the union bound technique and the obtained reliability metric.
    \item Studying the effects of the impairments and other key parameters on the PED, via numerical results, and providing insights into the performance of both systems in practice.
\end{itemize}

Next, section II introduces the communication model of the RIS-assisted SSK and RIS-assisted SSK-RPM with a sub-optimal greedy detector. In section III, we derive the closed-form expressions for the PED of the RIS-assisted SSK and the RIS-assisted SSK-RPM systems. The numerical results are presented in section IV, and section V concludes the paper.
\subsubsection*{Notations}
${\mathcal{CN}} \left(\mu,\sigma^2 \right)$ and ${\mathcal{N}} \left(\mu,\sigma^2 \right)$ respectively represent the distribution of a complex Gaussian and a real Gaussian random variable, with mean $\mu$ and variance $\sigma^2$. The notation $\left| \cdot \right|$ indicates the modulus operator, $\undb{E} \left[ \cdot \right]$ outputs the statistical expectation, $\Re\left\{ \cdot \right\}$ and $\Im \left\{ \cdot \right\}$ denote the real-part and imaginary-part operators, respectively, $\Gamma(\cdot)$ denotes the Gamma function, and $\jmath = \sqrt{-1}$.
\section{System and Channel Models}
\begin{figure}[!t]
    \centering
    \includegraphics[height=4.8cm, width=8cm]{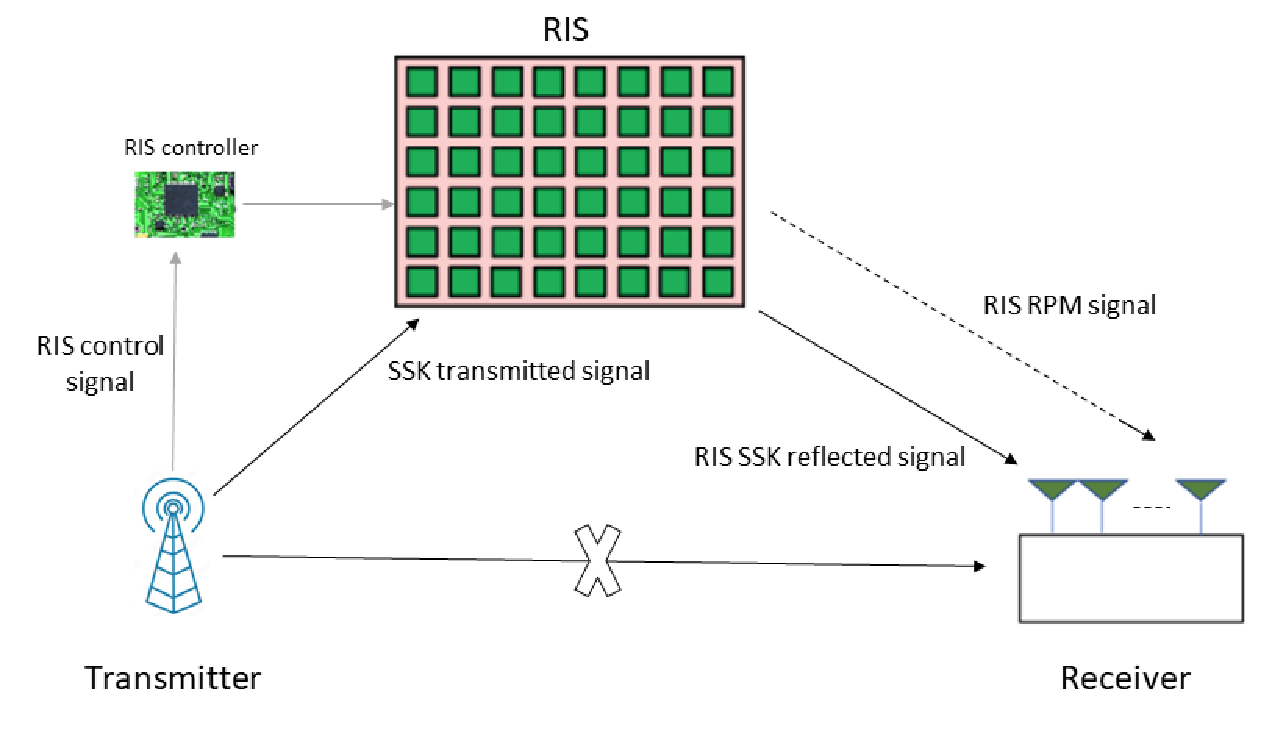}
    \caption{Illustration of the RIS-assisted SSK and RIS-assisted SSK-RPM communication models.}
    \label{figa}
\end{figure}

The proposed RIS-assisted communication model, as depicted in Fig. \ref{figa}, encompasses a transmission block comprised of a single antenna, and a RIS with $N$ regularly arranged reconfigurable elements. The RIS is considered to be placed close to the transmitter \cite{MaBhPa:14}, thus negating the effect of fading between them. The receiver, which is assumed to lie in the far field of the transmission block, consists of $N_R$ receive diversity branches, thus establishing a SIMO wireless communication set-up. The direct link from the transmitter to the receiver is assumed to be blocked.

The transmitter employs SSK-based or RPM-SSK-based modulation, and the receiver utilizes non-coherent greedy detection for determining the receive diversity branch with the maximum received energy.
Perfect CSI at the transmitter is assumed in both communication models, and the same is shared with the RIS through a controller in the transmission block.

The magnitude of the complex-valued channel gains between the RIS and the receive antennas are modeled to follow independent and identically distributed (i.i.d.) Nakagami-$m$ distributions \cite{zxc}. We express the channel gain between the $u$-th reconfigurable element of the RIS and the $w$-th diversity branch of the receiver in terms of its magnitude and phase as $h_{u,w}=\beta_{u,w} \exp \left\{- \jmath \theta_{u,w} \right\}$, $ w= 1, \ldots , N_{R}$, and $ u=1, \ldots, N$. Thus, the probability density functions (PDFs) of the $h_{u,w}$'s can be expressed as \cite{DaMaMo:16}
\beqarr
f_{\beta_{u,w},\theta_{u,w}} \left( \beta, \theta \right)
\! \! \! \! &=& \! \! \! \!
\frac{m^{m} \lvert\sin \theta \cos \theta \rvert^{m - 1} r^{2m - 1}
e^{-m \beta^2/\Omega}}
{\Omega^{m} \Gamma \left(\frac{\left(1 + p \right)m}{2} \right)
\Gamma \left(\frac{\left(1 - p\right)m}{2} \right)
\lvert \tan \theta \rvert^{pm}} \, , \nn \\
&& \hspace{-2.3cm}
w=1, \ldots, N_{R}, u= 1, \ldots ,N,
\beta>0, -\pi \leq \theta < \pi ,
\label{eqat}
\eeqarr
where $m$ denotes the shape parameter, $p$ is the power balance factor, and $\undb{E}\left[ \beta_{u,w}^2 \right] = \Omega$. The expression in Eq. (\ref{eqat}) can be formulated in terms of the PDFs of the in-phase and the quadrature-phase components of the channel gains by considering
\beq
h_{u,w}=X_{u,w}+\jmath Y_{u,w}=\beta_{u,w}\exp(-j\theta_{u,w}),
\eeq
with the distributions of $X_{u,w}$ and $Y_{u,w}$ given by
\beqarr
f_{X_{u,w}}(x) \! \! \! \! &=& \! \! \! \!
\left(\frac{m}{\Omega}\right)^{\frac{\left(1+p \right)m}{2}}
\frac{\lvert x\rvert^{\left(1+p \right)m - 1}}
{\Gamma\left(\frac{\left(1+p \right)m}{2}\right)}
\exp\left\{-{\frac{mx^{2}}{\Omega}}\right\} \, , \nn \\
&& \hspace{3.5cm} -\infty < x < \infty \, , \nn \\
f_{Y_{u,w}}(y) \! \! \! \! &=& \! \! \! \!
\left(\frac{m}{\Omega}\right)^{\frac{\left(1-p \right)m}{2}}
\frac{\lvert y\rvert^{\left(1-p \right)m - 1}}
{\Gamma\left(\frac{\left(1-p \right)m}{2}\right)}
\exp\left\{-{\frac{my^{2}} {\Omega}}\right\}\, , \nn \\
&& \hspace{3.5cm} -\infty < y < \infty \, .
\label{eqxy}
\eeqarr
The PDFs in (\ref{eqxy}) are even functions of $x$ and $y$, implying that $\undb{E}\left[ X_{u,w} \right] = \undb{E}\left[ Y_{u,w} \right]=0$, which further results in $\undb{E} \left[ h_{u,w} \right]=0$. Besides, the PDFs of the magnitude of the complex-valued channel gains can be expressed as \cite{dsgba}
\beq
f_{\beta_{u,w}}(r) = \frac{2m^{m}r^{2m - 1}}{\Omega^{m}\Gamma(m)}
\exp \left\{-\frac{mr^{2}}{\Omega} \right\} \, ,
\eeq
which further results in the first and second central moments of the magnitude of the complex-valued channel gains to be obtained as
\beqarr
\undb{E} \left[ \beta_{u,w} \right] \! \! \! \! &=& \! \! \! \!
\sqrt{\frac{\Omega}{m}}
\frac{\Gamma \left(m+\frac{1}{2} \right)}
{\Gamma \left( m \right)} \, , \nn \\
\undb{E} \left[ \left( \beta_{u,w} - \undb{E} \left[ \beta_{u,w} \right] \right)^2 \right]
\! \! \! \! &=& \! \! \! \!
\Omega \left(1- \frac{1}{m}
\left( \frac{\Gamma \left(m+\frac{1}{2} \right)}
{\Gamma \left( m \right)} \right)^2 \right) \, , \nn \\
&& \hspace{-1.5cm}
w=1, \ldots, N_{R}, u= 1, \ldots ,N, \beta>0 \, .
\label{eq:nakpar}
\eeqarr

Next, we describe the RIS-assisted SSK and RIS-assisted SSK-RPM systems considered in this study.
\subsection{RIS-Assisted SSK}
In the RIS-assisted SSK communication model, the SSK-modulated signal is intelligently reflected by the RIS to maximize the received signal-to-noise ratio (SNR) at the designated receive diversity branch. Maximum energy and, thus, maximum instantaneous SNR, observed at a particular receive diversity branch by the receiver's detector forms the basis for the selection of the intended branch. The incoming bit stream contains $\log_{2}{N_{R}}$ overhead bits dedicated to the identification of the index of the target receive diversity branch, and the rest of the bits are reserved for data transmission.

Due to the non-ideal nature of the transceiver pair, distortion noise will affect the received signal. A generalized model for the hardware impairment \cite{henry} is adopted. Consequently, the received signal at the receive antenna branch indexed by $w$ can be expressed as
\beq
z_w = \left(\sqrt{E_s}+ q_t\right) \left[ \sum_{u=1}^{N} h_{u,w} \exp \left\{\jmath \phi_u \right\} \right] + q_r +  n_w \, ,
\label{eq2}
\eeq
where $E_s$ is the transmit signal energy of the unmodulated carrier, $\phi_u$ is the phase shift introduced by the $u$-th RIS element, and $n_w \sim {\mathcal{CN}} \left(0, N_{0}\right)$ is the additive white Gaussian noise (AWGN) sample at the $w$-th receive antenna, which is assumed to be independent of $h_{u,w}$ and $\phi_{u}$ for $u \in \left\{1,\ldots,N\right\}$.

The variation from the ideal model is reflected by the presence of the distortion noise terms $q_t$ and $q_r$, which are used to model the residual impairments at the transmitter and the receiver of the communication system (RIS-assisted SSK or RIS-assisted SSK-RPM), respectively. In \cite{sbva213}, the authors suggested modeling these distortions using complex Gaussian random variables with zero mean and variances $k_{t}^2E_s$ and $k_{r}^2E_s\left|\sum_{u=1}^{N} h_{u,w}\right|^2$, respectively, where $k_t$ and $k_r$ are the hardware impairment levels determined using the ratio of the magnitude of the average distortion to the magnitude of the average signal or the error vector magnitude (EVM) metric. Without loss of generality, the distortions can be modeled using an aggregate distortion level $k$ such that $k = \sqrt{k_{t}^2 + k_{r}^2 }$. Thus, the expression \ref{eq2} of the received symbol can be re-formulated as
\beq
z_w = \left(\sqrt{E_s}+ q\right) \left[ \sum_{u=1}^{N} h_{u,w} \exp \left\{\jmath \phi_u \right\} \right]  +  n_w \, ,
\label{eq3}
\eeq
where $q \sim {\mathcal{CN}} \left(0, k^2 E_s \right)$ can be characterized as an independent distortion noise. For an ideal transceiver pair, we have $k=0$.
\subsection{RIS-Assisted SSK-RPM}
Inspired by the concept of RIS-IM \cite{MaBhPa:14}, and the idea of utilizing RIS as a transmitter \cite{sbz} for data sensed in the RIS-controller embedded with sensors to collect environmental data modulated in terms of discrete phase shifts governed by $M$-PSK modulation scheme, and appended to the SSK reflected signal from the transmitter.
Thus, the RIS plays the dual role of an intelligent reflector of the incoming SSK modulated signal, and a transmitter to modulate data in terms of discrete phase shifts and appending the sensed data with the SSK reflected signal.

Given that the transmitter and receiver are considered to be non-ideal, the received signal, affected by distortion noise, can be expressed as
\beq
z_{w,rp}\!\!=\!\! \left(\!\!\sqrt{E_s}\! +\! q_t\!\right)\!\!\! \left[ \sum_{u=1}^{N}\! h_{u,w} \exp \left\{\jmath \phi_u \right\} \!\!\right]\!\!\exp \left\{\jmath \psi_n \right\}
+ q_r+ n_{w} \, ,  \\
\label{eq6}
\eeq
where $\psi_n, \forall n=\left\{1, \ldots, M\right\}$, denotes the discrete phase selected from the set of equiprobable $M$-ary RPM constellation. To further simplify, we can use the aggregate distortion level $q$ and reformulate Eq. (\ref{eq6}) as
\beq
z_{w,rp} = \left(\sqrt{E_s} + q\right)\!\!\! \left[ \sum_{u=1}^{N} \! h_{u,w} \exp \left\{\jmath \phi_u \right\}\!\! \right]\exp \left\{\jmath \psi_n \right\}
+ n_{w} \, .
\label{eq7}
\eeq
\subsection{Greedy Detector}
The receivers of the RIS-assisted SSK and RIS-assisted SSK-RPM systems utilize a non-coherent greedy detector. The latter ascertains the index of the receive diversity branch associated with the maximum instantaneous SNR across all the diversity branches without requiring knowledge of the CSI at the receiving end. Consequently, the decision rule employed to select the index of the target receive branch is given by
\beq
\hat{p}= \arg \max_p \left| z_p \right|^2 \, , \ 1\leq p \leq N_R \, .
\label{eq13}
\eeq
\section{Performance Analysis}
Considering the models for the RIS-assisted SSK and RIS-assisted SSK-RPM communication systems with greedy-based detection, in this section, we present the analytical framework for obtaining closed-form expressions for the system performance metric PED and its specific case PPED for the case of receiver with two receive diversity branches subject to hardware impairments.
\subsection{PED of the RIS-Aided SSK System}
Let $w$ and $\hat{w}$ be the indices of the target branch and a non-target antenna branch, respectively. Utilizing (\ref{eq3}) and (\ref{eq13}), we can express the pairwise PED (PPED) as
\beqarr
&& \hspace{-1.5cm }\Pr \left\{ \left| z_w \right|^2 < \left| z_{\hat{w}}  \right|^2 \right\}  \nn \\ &&
\hspace{-0.8cm}
= \Pr \left\{ \left| \left( \sum_{u=1}^{N} h_{u,w} e^{\jmath \phi_{u}} \right) \left(\sqrt{E_s}+q\right) + n_w \right|^2 \right. \nn \\
&&  < \left. \left|\left(\sum_{u=1}^{N} h_{\hat{u},w} e^{\jmath \phi_{u}} \right) \left(\sqrt{E_s}+q\right) + n_{\hat{w}} \right|^2 \right\}.
\label{qw2}
\eeqarr
The expression in Eq. (\ref{qw2}) can be minimized by maximizing the energy of the signal at the target antenna branch, which can be achieved by substituting $\phi_u$ with $\theta_{u,w}$ for $u = 1,\cdots,N$,  considering CSI availability at the transmitter.\footnote{We assume that the transmitter can share the instantaneous CSI data with the RIS controller in real-time such that we can steer the signal accordingly.} Thus, Eq. (\ref{qw2}) can be simplified as
\beqarr
&& \hspace{-1.5cm} \Pr \left\{ \left| z_w \right|^2 < \left| z_{\hat{w}} \right|^2 \right\} \nn \\
&&
\hspace{-1.0cm}
= \Pr \left\{ \left| \left( \sum_{u=1}^{N} \alpha_{u,w}  \right) \left(\sqrt{E_s}+q\right) + n_w \right|^2 \right. \nn \\
&& \hspace{-0.55cm}
< \left. \left| \left(\sqrt{E_s}+q\right) \sum_{u=1}^{N} \alpha_{\hat{u},w}
e^{ \jmath \left( \theta_{u,w} - \theta_{\hat{u},w} \right)}
\! + \! n_{\hat{w}} \right|^2 \! \right\} .
\label{qw3as}
\eeqarr

{\em Theorem 1}: The closed-form expression for the PPED of the RIS-assisted SSK system with hardware impairment is given by Eq. (\ref{titanic}) on the next page, where $\Gamma_{av} \dn E_s\Omega/N_0$ denotes the average SNR of the system.

{\em Proof}: The proof is presented in Appendix A. \hfill $\blacksquare$

\begin{figure*}
\beqarr
\Pr \left \{{ X_{sk} < Y_{sk} }\right \} &=&
\frac{\left(N\Gamma_{av}+Nk^2\Gamma_{av}+1\right)}{\left(  N\Gamma_{av} +Nk^2\Gamma_{av}\left(\!2+\!\frac{1}{m}\left(\frac{\Gamma(m+\frac{1}{2})}{\Gamma(m)}\right)^2\right)+ 2\right)^{\frac{1}{2}}}
\nn \\ && \hspace{-0.5cm} \times
\frac{\exp \left\{-
\frac{ \frac{N^2 \Gamma_{av}}{m}  \frac{\Gamma^2\left({m+\frac{1}{2}}\right)}{\Gamma^2\left({m}\right)}}
{\left(  N\Gamma_{av} +Nk^2\Gamma_{av}\left(\!2+\!\frac{1}{m}\left(\frac{\Gamma(m+\frac{1}{2})}{\Gamma(m)}\right)^2\right)+ 2 +  2\left(N\Gamma_{av}\left(\!1-\!\frac{1}{m}\left(\frac{\Gamma(m+\frac{1}{2})}{\Gamma(m)}\right)^2\right)\right)\right)}\right\}}
{\left(  N\Gamma_{av} +Nk^2\Gamma_{av}\left(\!2+\!\frac{1}{m}\left(\frac{\Gamma(m+\frac{1}{2})}{\Gamma(m)}\right)^2\right)+ 2 +  2\left(N\Gamma_{av}\left(\!1-\!\frac{1}{m}\left(\frac{\Gamma(m+\frac{1}{2})}{\Gamma(m)}\right)^2\right)\right)\right)^{\frac{1}{2}}
}.
\label{titanic}
\eeqarr
\noindent\rule{\textwidth}{.5pt}
\end{figure*}

To this point, we derived the PPED reliability metric considering two receive antenna branches. To extend our analysis to a more general configuration with $N_{R} \left( >2 \right)$ receive diversity branches, we consider the random variables (RVs) \( Y_{1,_{sk}}, \ldots, Y_{L,_{sk}} \) denoting the instantaneous signal energies at the non-target receive diversity antennas, with \( L = N_{R} - 1 \). Exploiting these i.i.d. variables, Eq. (\ref{ab1}) is modified to derive the PED as
\beqarr
P_{e,sk} \! \! \! \! &=&  1 - \Pr \left\{Y_{1,_{sk}}, \ldots ,Y_{L,_{sk}} < X \right\}  \nn \\
&=& \int\limits_{-\infty}^{\infty}\left(\prod \limits_{i=1}^{L} F_{Y_{i,sk}} \left(x \right) \right)
f_{X_{sk}} \left(x\right) \textnormal{d}x .
\label{eq26}
\eeqarr
Utilizing the statistics obtained in Eq. (\ref{qw5a}) and leveraging the i.i.d. property of the RVs, we have
\beqarr
&& \hspace{-1.2cm}
\prod \limits_{i=1}^{L} F_{Y_{i,sk}} (x)
= \left(1 - \exp\left\{- \frac{x}{a}\right\} \right)^{L} \nn \\
&& \quad
= \sum \limits_{r=1}^{L} (-1)^{r-1}
\left(\begin{array}{c} L\\ r \end{array}\right)
\exp \left\{-\frac{rx}{a} \right\}, x \geq 0 \, ,
\label{eq27}
\eeqarr
where the RV $a$ is as defined in Eq. (\ref{eq18}) (cf. Appendix A).

Substituting (\ref{eq27}) into (\ref{eq26}), and subsequently performing algebraic simplifications, we obtain the generalized PED for the RIS-assisted SSK communication system with non-ideal transceivers as shown in Eq.  (\ref{britanic}) on the next page.

\begin{figure*}
\beqarr
\hspace{0.5cm} P_{e,sk} &=& \left(N\Gamma_{av}+Nk^2\Gamma_{av}+1\right) \sum \limits _{r=1}^{L}
\frac{(-1)^{r-1} \left ({{\begin{array}{c} L\\ r\end{array}}}\right)}{\left(  N \Gamma_{av} + r+1 +  Nk^2\Gamma_{av}\left(1+r\left(1+ \frac{1}{m}\left(\frac{\Gamma(m+\frac{1}{2})}{\Gamma(m)}\right)^2\right)\right)
\right)^{\frac{1}{2}}} \nn \\ \hspace{-5.4cm} &\times&
\frac{ \exp \left\{-\frac{   \frac{rN^2\Gamma_{av}}{m} \frac{\Gamma^2\left({m+\frac{1}{2}}\right)}{\Gamma^2\left({m}\right)}}
{\left(  N \Gamma_{av} + r+1 +  Nk^2\Gamma_{av}\left(1+r\left(1+ \frac{1}{m}\left(\frac{\Gamma(m+\frac{1}{2})}{\Gamma(m)}\right)^2\right)\right)
+2r\left(N\Gamma_{av}\left(\!1-\!\frac{1}{m}\left(\frac{\Gamma(m+\frac{1}{2})}{\Gamma(m)}\right)^2\right)\right)\right)}\right\}}
{\left(  N \Gamma_{av} + r+1 +  Nk^2\Gamma_{av}\left(1+r\left(1+ \frac{1}{m}\left(\frac{\Gamma(m+\frac{1}{2})}{\Gamma(m)}\right)^2\right)\right)
+2r\left(N\Gamma_{av}\left(\!1-\!\frac{1}{m}\left(\frac{\Gamma(m+\frac{1}{2})}{\Gamma(m)}\right)^2\right)\right)\right)^{\frac{1}{2}}
}.
\label{britanic}
\eeqarr
\noindent\rule{\textwidth}{.5pt}
\end{figure*}
\subsubsection{Asymptotic Analysis}
For scenarios pertaining to high values of the average SNR $\Gamma_{av}$, the PED in Eq. $\left(\ref{britanic}\right)$ will simplify to the expression shown in Eq. $\left(\ref{hfda}\right)$. On the other hand, for the case of low $\Gamma_{av}$, Eq. $\left(\ref{britanic}\right)$ simplifies to
\beq
P_{e,sk_{\Gamma_{av}}\! \ll 1}=\! \! \sum \limits _{r=1}^{L} (-1)^{r-1} \left({{\!\!\!\begin{array}{c} L\\ r \end{array}}}\!\!\!\right) \frac{\text{exp} \left\{-\frac{rN^2\Gamma_{av}}{ m\left(r+1\right)}\frac{\Gamma^2\left({m+\frac{1}{2}}\right)}{\Gamma^2\left({m}\right)}\right\}}{\left(r+1\right)} .
\label{eq30}
\eeq
Additionally, for $\Gamma_{av}=0$, Eq. $\left(\ref{britanic}\right)$ reduces to
\beq
P_{e,sk_{\Gamma_{av}=0}\!}=\!  \sum \limits _{r=1}^{L} \!(-1)^{r-1} \left ({{\!\!\!\begin{array}{c} L\\ r \end{array}}}\!\!\!\right) \frac{1}{\left(r+1\right)} \stackrel{(b)}{=} \frac{L}{L+1}.
\label{eq23ty}
\eeq
The proof of step $(b)$ in Eq. (\ref{eq23ty}) is provided in Appendix B.

\begin{figure*}
\beqarr
 P_{e,sk_{\Gamma_{av}}\! \gg 1} \!\!\!\!\! &=& \!\!\!\!\!
\frac{\left(1+k^2\right)\sum \limits _{r=1}^{L} (-1)^{r-1} \left ({{\begin{array}{c} L\\ r \end{array}}}\right) \exp \left\{-\frac{  \frac{rN\Gamma^2\left({m+\frac{1}{2}}\right)}{m\Gamma^2\left({m}\right)}}{\left(1+k^2\left(1+r\left(\!1+\!\frac{1}{m}\left(\frac{\Gamma(m+\frac{1}{2})}{\Gamma(m)}\right)^2\right)\right) +  2r\left(\!1-\!\frac{1}{m}\left(\frac{\Gamma(m+\frac{1}{2})}{\Gamma(m)}\right)^2\right)\right)}\right\}}{\left(\!\!1+k^2\left(\!\!1+r\left(\!1+\!\frac{1}{m}\left(\frac{\Gamma(m+\frac{1}{2})}{\Gamma(m)}\right)^2\right)\right)\!\! + \!  2r\left(\!1-\!\frac{1}{m}\left(\frac{\Gamma(m+\frac{1}{2})}{\Gamma(m)}\right)^2\right)\right)^{\frac{1}{2}}
\!\!\!
\left(\!\!1+k^2\!\left(\!\!1+r\left(\!1+\!\frac{1}{m}\left(\frac{\Gamma(m+\frac{1}{2})}{\Gamma(m)}\right)^2\right)\right)\right)^\frac{1}{2}}.
\nn \\
\label{hfda}
\eeqarr
\noindent\rule{\textwidth}{.5pt}
\end{figure*}

\textit{Remark 1:} According to the expressions in (\ref{eq30}), (\ref{eq23ty}), and (\ref{hfda}), the distortion noise in the non-ideal transceivers leads to a destructive effect at high average SNRs and has minimal to no impact in the lower range values of the average SNR. Thus, for the system with the greedy detector, the error floor of the reliability metric PED at higher values of the average SNR $\Gamma_{av}$ only gets affected according to the aggregate distortion level $k$.
\subsection{PED of the RIS-Assisted SSK-RPM System}
For the RIS-assisted SSK-RPM system, we consider the target and a non-target antenna to be indexed by $w,rp$ and $\hat{w},rp$, respectively. Thus, using Eqs. (\ref{eq7}) and (\ref{eq13}), we can compute the PPED as
\beqarr
&& \!\!\!\!\!\!\!\!\!\!\!\!\!\!\!\!\!\!\!
{\Pr \left \{{ \left |{z_{w,rp} }\right |^{2} < \left |{ z_{\hat{w},rp} }\right |^{2} }\right \}} \nn \\
&& \!\!\!\!\!\!\!\!\!\!\!\!\!\!
= {\Pr \Biggl \{{\left|{ \left(\sqrt{E_s}+q\right)\left ({\sum \limits _{u=1}^{N} h_{u,w} e^{\jmath \phi _{u}} }\right)e^{\jmath \psi _{n}}  + n_{w} }\right|^{2}}} \nn \\
&& {<}
{{ \left |{\left(\sqrt{E_s}+q\right) \left({{\sum \limits _{u=1}^{N} h_{\hat{u},w} e^{\jmath \phi_{u}}} }\right)e^{\jmath \psi _{n}}  + n_{\hat w} }\right |^{2} }\Biggr \},}
\label{qw3a}
\eeqarr
where $\psi_n$ is the additional information modulated by the RIS and appended to the SSK signal. For maximizing the received signal at the intended branch or minimizing the above expression, we set $\phi_u = \theta_{u,w}$ for $u=1,\ldots,N$. Thus, Eq. (\ref{qw3a}) can be simplified as follows:
\beqarr
&& \!\!\!\!\!\!\!\!\!\!\!
{\Pr \left \{{ \left |{z_{w,rp} }\right |^{2} < \left |{ z_{\hat{w},rp} }\right |^{2} }\right \}} \nn \\
&& \!\!\!\!\!\!\!\!
= {\Pr \Biggl \{{\left|{ \left(\sqrt{E_s}+q\right)\left ({\sum \limits _{u=1}^{N} \alpha_{u,w}  }\right)e^{\jmath \psi _{n}}  + n_{w} }\right|^{2}}} \nn \\
&& \!\!\!\!\!\!
{<}
{ \left |{\left(\sqrt{E_s}+q\right) \left({{\sum \limits _{u=1}^{N} \alpha_{\hat{u},w} e^{\jmath\left( \theta_{u,w} - \theta_{\hat u,w}\right)}} }\right)e^{\jmath \psi _{n}}  + n_{\hat w} }\right |^{2} }\Biggr \} . \nn \\
\label{eq33}
\eeqarr

{\em Theorem 2}: The closed-form expression for the PPED of the RIS-assisted SSK-RPM system is given by Eq. (\ref{purisf}).

{\em Proof}: The proof is presented in Appendix C. \hfill $\blacksquare$

\begin{figure*}
\beqarr
&& \! \! \! \! \! \! \! \! \! \! \!
\! \! \!
\Pr \left\{ \left. \! X_{rp} < Y_{rp} \right| \psi_n \right\}
= \left( N\Gamma_{av}+Nk^2\Gamma_{av}+1\right)
\nn \\
&& \times
\frac{
\exp\left\{-\frac{\frac{N^2 \Gamma_{av} \left(\sin \psi_n\right)^2
\Gamma^2\left(m+\frac{1}{2} \right)}{m\Gamma^2\left({m}\right)}}
{N\Gamma_{av} + Nk^2\Gamma_{av} + 2+ 2 \left( N\Gamma_{av}
\left(\sin\psi_n\right)^2
\left(1-\frac{\Gamma^2\left(m+\frac{1}{2}\right)}{m\Gamma^2
\left(m\right)}\right)\right)+Nk^2\Gamma_{av}\left(1+\frac{\Gamma^2\left(m+\frac{1}{2}\right)}{m\Gamma^2
\left(m\right)}\right)}
\right\}}
{\left( N\Gamma_{av} + Nk^2\Gamma_{av} + 2+ 2 \left( N\Gamma_{av}
\left(\sin \psi_n\right)^2
\left(1-\frac{\Gamma^2\left(m+\frac{1}{2}\right)}{m\Gamma^2
\left(m\right)}\right)\right)+Nk^2\Gamma_{av}\left(1+\frac{\Gamma^2\left(m+\frac{1}{2}\right)}{m\Gamma^2
\left(m\right)}\right)\right)^{\frac{1}{2}} \! \!
} \nn \\
&& \hspace{1.6cm} \times
\frac{\exp\left\{-\frac{\frac{N^2 \Gamma_{av} \left(\cos \psi_n\right)^2
\Gamma^2\left(m+\frac{1}{2} \right)}{m\Gamma^2\left({m}\right)}}
{N\Gamma_{av} + Nk^2\Gamma_{av} + 2+ 2 \left( N\Gamma_{av}
\left(\cos \psi_n\right)^2
\left(1-\frac{\Gamma^2\left(m+\frac{1}{2}\right)}{m\Gamma^2
\left(m\right)}\right)\right)+Nk^2\Gamma_{av}\left(1+\frac{\Gamma^2\left(m+\frac{1}{2}\right)}{m\Gamma^2
\left(m\right)}\right)}\right\}}{\left(N\Gamma_{av} + Nk^2\Gamma_{av} + 2+ 2 \left( N\Gamma_{av}
\left(\cos \psi_n\right)^2
\left(1-\frac{\Gamma^2\left(m+\frac{1}{2}\right)}{m\Gamma^2
\left(m\right)}\right)\right)+Nk^2\Gamma_{av}\left(1+\frac{\Gamma^2\left(m+\frac{1}{2}\right)}{m\Gamma^2
\left(m\right)}\right)\right)^\frac{1}{2}}.
\label{purisf}
\eeqarr
\noindent\rule{\textwidth}{.5pt}
\end{figure*}

Extending the analysis to a more general SIMO communication scenario with $N_{R}$ receive antennas, as done in the case of the RIS-assisted SSK system, we exploit the i.i.d. nature of the RVs denoting the energies at the non-target antennas, i.e., $Y_{1_{rp}}, \ldots, Y_{{L}_{rp}}$, and accordingly find the conditional PED of the target antenna of the RIS-assisted SSK-RPM system as
\beqarr
P_{c,rp}|\psi_{n} \! \! \! \! &=& \! \! \! \!
1 - \Pr \left\{Y_{1_{rp}}, \ldots ,Y_{{L}_{rp}} < X_{rp}|\psi_n \right\} \nn \\
&=& \! \! \! \! \int\limits_{-\infty}^{\infty}\left(\prod \limits_{i=1}^{L}
F_{Y_{i_{rp}}} \left(x \right) \right) f_{X_{rp}} \left(x\right) \textnormal{d}x .
\label{eq44}
\eeqarr
where $\prod \limits_{i=1}^{L}F_{Y_{i_{rp}}} \left(x \right)$ is given by
\beqarr
&& \! \! \! \! \! \! \! \! \! \! \! \! \! \! \! \!
\! \! \! \! \! \!
\prod \limits_{i=1}^{L} F_{Y_{i_{rp}}} (x)
= \left(1 - \exp\left\{- \frac{x}{a}\right\} \right)^{L} \nn \\
&&
= \sum \limits_{r=1}^{L} (-1)^{r-1}
\left( \! \! \begin{array}{c} L\\ r \end{array} \! \! \right)
\exp \left\{-\frac{rx}{a} \right\} ,
x \geq 0,
\label{eq45}
\eeqarr
with the RV $a$ already defined in Eq. (\ref{eq18}) of Appendix A. Therefore, the final unconditioned expression for the general case with $N_R$ receive diversity antennas is as shown in Eq. (\ref{eq47a}) at the top of the next page.

\begin{figure*}
\beqarr
&& \! \! \! \! \! \! \! \! \! \! \!
\! \! \! \! \! \! \! \! \! \! \! \!
P_{e,rp}
= \frac{\left( N\Gamma_{av}+Nk^2\Gamma_{av}+1\right)}{M} \nn \\
&& \times
\sum_{n=1}^{M}\sum \limits_{r=1}^{L}
\frac{ (-1)^{r}
\left( \! \! \begin{array}{c} L\\ r \end{array} \! \! \right)
\exp\left\{-\frac{\frac{rN^2 \Gamma_{av} \left(\sin \psi_n\right)^2
\Gamma^2\left(m+\frac{1}{2} \right)}{m\Gamma^2\left({m}\right)}}
{r+1 + N\Gamma_{av} \left(1+2r
\left(\sin\psi_n\right)^2
\left(1-\frac{\Gamma^2\left(m+\frac{1}{2}\right)}{m\Gamma^2
\left(m\right)}\right)\right)+Nk^2\Gamma_{av}\left(1+r\left(1+\frac{\Gamma^2\left(m+\frac{1}{2}\right)}{m\Gamma^2
\left(m\right)}\right)\right)}
\right\}}
{\left(r+1 + N\Gamma_{av} \left(1+2r
\left(\sin\psi_n\right)^2
\left(1-\frac{\Gamma^2\left(m+\frac{1}{2}\right)}{m\Gamma^2
\left(m\right)}\right)\right)+Nk^2\Gamma_{av}\left(1+r\left(1+\frac{\Gamma^2\left(m+\frac{1}{2}\right)}{m\Gamma^2
\left(m\right)}\right)\right)\right)^{\frac{1}{2}} \! \!
} \nn \\
&& \hspace{1.2cm} \times
\frac{\exp\left\{-\frac{\frac{rN^2 \Gamma_{av} \left(\cos \psi_n\right)^2
\Gamma^2\left(m+\frac{1}{2} \right)}{m\Gamma^2\left({m}\right)}}
{r+1 + N\Gamma_{av} \left(1+2r
\left(\cos\psi_n\right)^2
\left(1-\frac{\Gamma^2\left(m+\frac{1}{2}\right)}{m\Gamma^2
\left(m\right)}\right)\right)+Nk^2\Gamma_{av}\left(1+r\left(1+\frac{\Gamma^2\left(m+\frac{1}{2}\right)}{m\Gamma^2
\left(m\right)}\right)\right)}\right\}}
{\left(r+1 + N\Gamma_{av} \left(1+2r
\left(\cos\psi_n\right)^2
\left(1-\frac{\Gamma^2\left(m+\frac{1}{2}\right)}{m\Gamma^2
\left(m\right)}\right)\right)+Nk^2\Gamma_{av}\left(1+r\left(1+\frac{\Gamma^2\left(m+\frac{1}{2}\right)}{m\Gamma^2
\left(m\right)}\right)\right)\right)^\frac{1}{2}}.
\label{eq47a}
\eeqarr
\noindent\rule{\textwidth}{.5pt}
\end{figure*}
\subsubsection{Asymptotic Analysis}
For scenarios with high values of the average SNR $\Gamma_{av}$, Eq. (\ref{eq47a}) will simplify to the expression shown in  (\ref{eq48}). Additionally, for scenarios characterized by low average SNR, Eq. (\ref{eq47a}) simplifies to
\beqarr
&& \! \! \! \! \! \! \! \! \! \! \!
\! \! \! \! \! \! \! \! \! \! \! \!
P_{e,{rp},\Gamma_{av} \ll 1}
= \frac{1}{M}\sum_{n=1}^{M}\sum \limits_{r=1}^{L}
\frac{ (-1)^{r}
}
{\left(r+1 \right) \! \!
}\left( \! \! \begin{array}{c} L\\ r \end{array} \! \! \right) \nn \\
&&  \times
\exp\left\{-\frac{rN^2 \Gamma_{av} \left(\sin \psi_n\right)^2
\Gamma^2\left(m+\frac{1}{2} \right)}{m\Gamma^2\left({m}\right)\left(r+1\right)}
\right\} \nn \\
&& \times
\exp\left\{-\frac{rN^2 \Gamma_{av} \left(\cos \psi_n\right)^2
\Gamma^2\left(m+\frac{1}{2} \right)}{m\Gamma^2\left({m}\right)\left(r+1\right)}
\right\} .
\label{eq49}
\eeqarr
Furthermore, when $\Gamma_{av}=0$, Eq. $\left(\ref{eq47a}\right)$ simplifies to
\beq
P_{e,{rp}, \Gamma_{av}=0}
= \sum\limits_{r=1}^{L} (-1)^{r-1}
\left ({{\!\!\begin{array}{c} L\\ r \end{array}}}\!\!\right) \frac{1}{\left(r+1\right)} \stackrel{(b)}{=} \frac{L}{L+1}.
\label{eq42da}
\eeq
The proof of step $(b)$ above is provided in Appendix B. \hfill $\blacksquare$

\begin{figure*}
\beqarr
&& \! \! \! \! \! \! \! \! \! \! \!
\! \! \! \! \! \! \! \! \! \! \! \!
P_{e,{rp}_{\Gamma_{av}} \gg 1}
= \frac{\left( 1+k^2\right)}{M}\sum_{n=1}^{M}\sum \limits_{r=1}^{L}
\frac{ (-1)^{r}
\left( \! \! \begin{array}{c} L\\ r \end{array} \! \! \right)
\exp\left\{-\frac{\frac{rN  \left(\sin \psi_n\right)^2
\Gamma^2\left(m+\frac{1}{2} \right)}{m\Gamma^2\left({m}\right)}}
{ \left(1+2r
\left(\sin\psi_n\right)^2
\left(1-\frac{\Gamma^2\left(m+\frac{1}{2}\right)}{m\Gamma^2
\left(m\right)}\right)\right)+k^2\left(1+r\left(1+\frac{\Gamma^2\left(m+\frac{1}{2}\right)}{m\Gamma^2
\left(m\right)}\right)\right)}
\right\}}
{\left(  \left(1+2r
\left(\sin\psi_n\right)^2
\left(1-\frac{\Gamma^2\left(m+\frac{1}{2}\right)}{m\Gamma^2
\left(m\right)}\right)\right)+k^2\left(1+r\left(1+\frac{\Gamma^2\left(m+\frac{1}{2}\right)}{m\Gamma^2
\left(m\right)}\right)\right)\right)^{\frac{1}{2}} \! \!
} \nn \\
&& \hspace{1.6cm} \times
\frac{\exp\left\{-\frac{\frac{rN  \left(\cos \psi_n\right)^2
\Gamma^2\left(m+\frac{1}{2} \right)}{m\Gamma^2\left({m}\right)}}
{  \left(1+2r
\left(\cos\psi_n\right)^2
\left(1-\frac{\Gamma^2\left(m+\frac{1}{2}\right)}{m\Gamma^2
\left(m\right)}\right)\right)+k^2\left(1+r\left(1+\frac{\Gamma^2\left(m+\frac{1}{2}\right)}{m\Gamma^2
\left(m\right)}\right)\right)}\right\}}
{\left(  \left(1+2r
\left(\cos\psi_n\right)^2
\left(1-\frac{\Gamma^2\left(m+\frac{1}{2}\right)}{m\Gamma^2
\left(m\right)}\right)\right)+k^2\left(1+r\left(1+\frac{\Gamma^2\left(m+\frac{1}{2}\right)}{m\Gamma^2
\left(m\right)}\right)\right)\right)^\frac{1}{2}}.
\label{eq48}
\eeqarr
\noindent\rule{\textwidth}{.5pt}
\end{figure*}

\textit{Remark 2:} From the above analysis (cf. eq. (\ref{eq42da})), it can be observed that the performance of the RIS-assisted SSK-RPM communication system at $\Gamma_{av}=0$ is independent of the constellation utilized for the RPM at the RIS or, conversely, that there is no added performance advantage from the utilization of the SSK-RPM modulation instead of the SSK modulation at this specified value of $\Gamma_{av}$.

\textit{Remark 3:} Based on (\ref{eq49}), (\ref{eq42da}) and (\ref{eq48}), we observe that the destructive effect of the distortion noise due to the non-ideal nature of the transceivers is limited to the ranges of high average SNR, which can be attributed to the structure of the detector used in the communication system under study.
\subsection{Error Rate Analysis}
By exploiting the fact that the PED was derived using the i.i.d. nature of the RVs characterizing the instantaneous signal energies at the diversity branches of the receiver, we can approximate the BER by using the union-bound technique. Thus, we have
\beq
\text{BER} \leq \frac{1}{\log_{2}N_R} \sum
P_{e} \, d^{\zeta}\left(w \to \hat{w}\right) =
\frac{N_R}{2} P_{e} \, ,
\label{eq55}
\eeq
where $P_{e}$ denotes either the PED of the SSK or the SSK-RPM modulation schemes, i.e., $P_{e,sk}$ or $P_{e,rp}$ depending on the scheme for which the BER is being computed, and $d^{\zeta}\left(\cdot\right)$ denotes the hamming distance between the two binary representations considered in the union bound computation.
\section{Numerical Results and Discussion}
In this section, we validate the proposed analytical framework with the aid of Monte-Carlo simulations over a total of $10^{9}$ channel realizations. We also illustrate the impact of different system parameters on the PED to provide further insights into the proposed RIS-assisted SSK and RIS-assisted SSK-RPM communication models.

\begin{figure}
    \centering
    \hspace*{-0.2cm}\includegraphics[height=7.0cm, width=9.0cm]{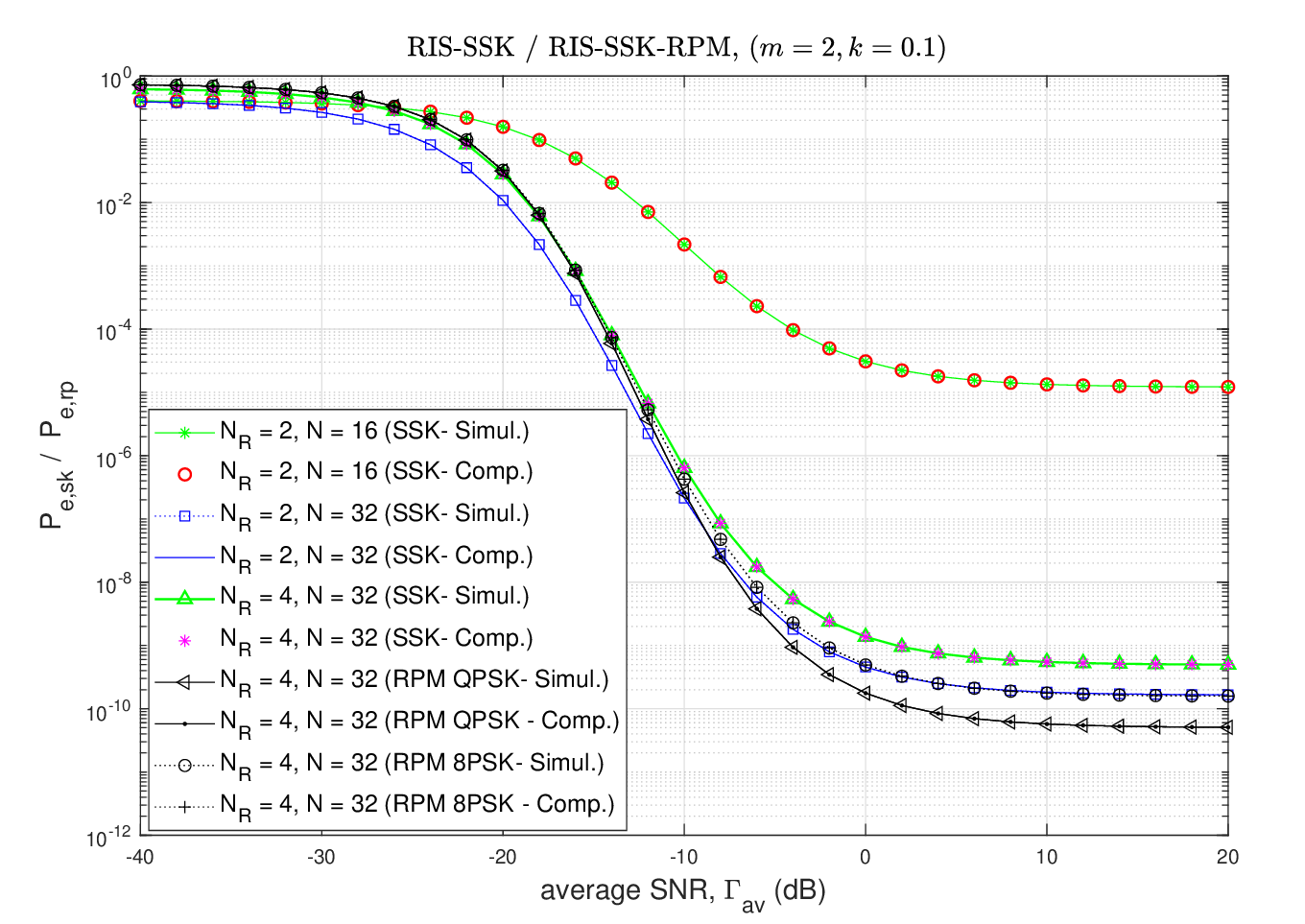}
    \caption{Simulation vs. computational results (cf. (28) for SSK and (48) for SSK-RPM) for impairment level $k$ = 0.1, of SSK and SSK-RPM (QPSK, 8PSK) modulation schemes.}
    \label{fig1}
\end{figure}

Fig. \ref{fig1} shows the plots of PED for the SSK and SSK-RPM modulation schemes versus the average SNR, for different values of $N_R$ and $N$, with the impairment level $k$ kept constant. The simulation and analytical plots coincide, validating the theoretical framework developed in this study. Furthermore, it is observed that $P_{e,sk}$ and  $P_{e,rp}$ decrease with increasing value of the average SNR $\Gamma_{av}$, and tend to saturate towards high values of $\Gamma_{av}$, which can be attributed to the structure of the greedy detector utilized at the receiver. Furthermore, $P_{e,sk}$ and $P_{e,rp}$ are concave for lowe values of $\Gamma_{av}$ and convex at high values of $\Gamma_{av}$ with a distinct intermediate point of inflection corresponding to a distinct $\Gamma_{av}$ for different values of $N$ and $N_R$.

\begin{figure*}
    \centering
    \includegraphics[height=7.4cm, width=18.6cm]{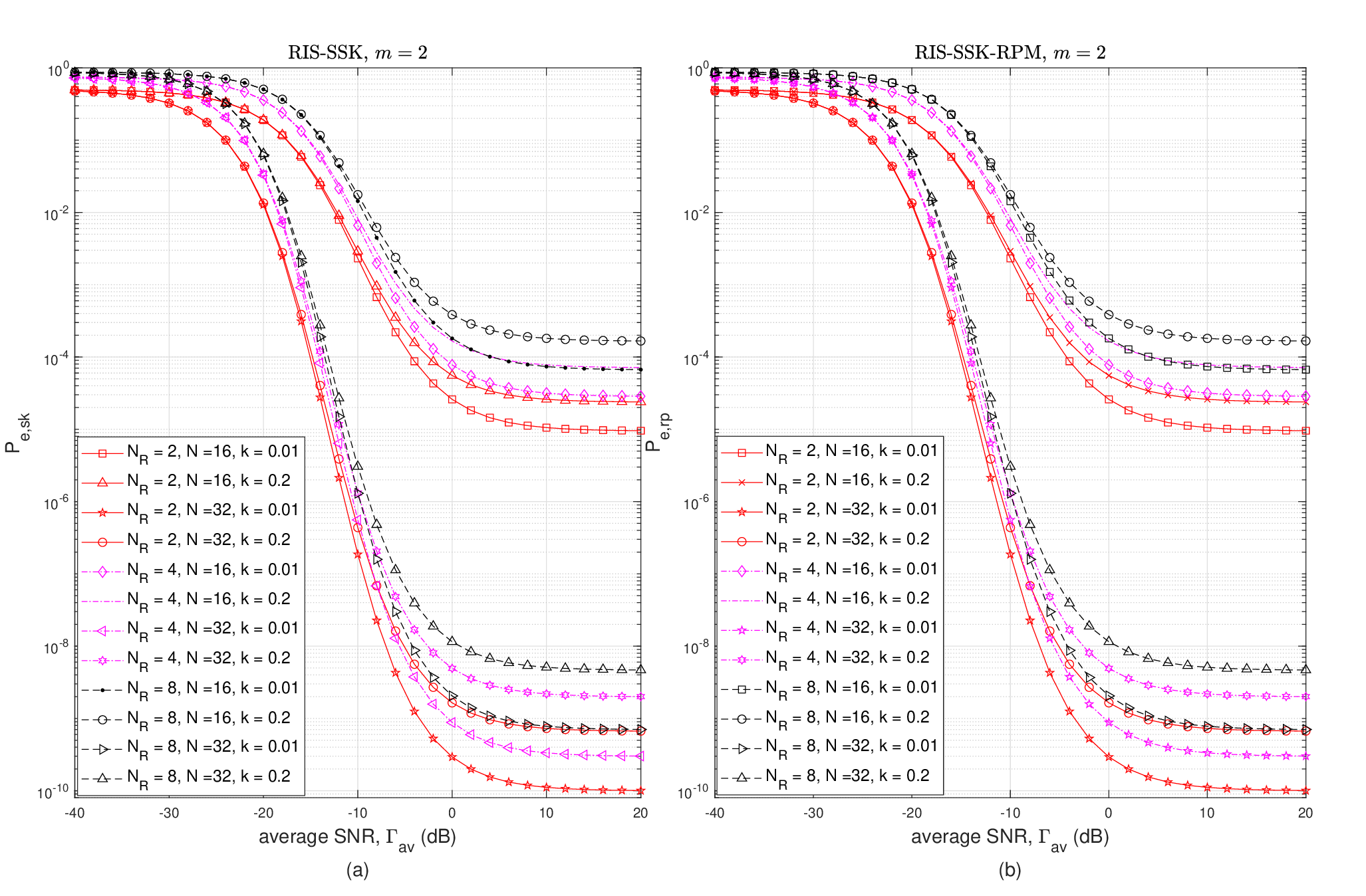}
    \caption{Computation plots (cf. (28) for SSK and (48) for RPM) for impairment level $k$ = 0.01, 0.2; (a) SSK modulation scheme, and (b) SSK-RPM (QPSK) modulation scheme.}
    \label{fig2}
\end{figure*}

Fig. \ref{fig2} illustrates the variation of the PED with $\Gamma_{av}$ for varying values of the system parameters $N$ and $N_R$, and different impairment levels $k$. We observe that for both constellations, i.e., RIS-assisted SSK and RIS-assisted SSK-RPM, the performance degrades as the impairment level $k$ increases from 0.01 to 0.2, with the effect being more prominent for higher values of $\Gamma_{av}$, more precisely on the floor value of PED at high $\Gamma_{av}$, due to the utilization of the greedy detector structure at the receiver. It should be noted that the effect of an increase in $k$ directly increases the distortion noise present in the system transceivers, affecting the signal quality and thus the system performance. Moreover, it can be observed that increasing the number of RIS elements $N$ and the number of receive antennas $N_R$ have conflicting effects on performance enhancements for the communication system under study. Additionally, we can also observe the prominent clustering effect based on the value of $N$. Furthermore, increasing $N_R$ and $k$ have reciprocal effects on the system performance enhancements, but are less prominent compared to the impact of parameter $N$, thus a slight shift in the inflection points of the PED curves can be expected accordingly. Conversely, it can be stated that the effect of hardware impairment with the greedy detector has minimal or no impact at lowe values of $\Gamma_{av}$. In contrast, the saturated value of the PED metric gets affected at high $\Gamma_{av}$ for fixed values of $N, N_R$ but different $k$.

\begin{figure}
    \centering
    \hspace*{-0.2cm}\includegraphics[height=7.4cm, width=9.5cm]{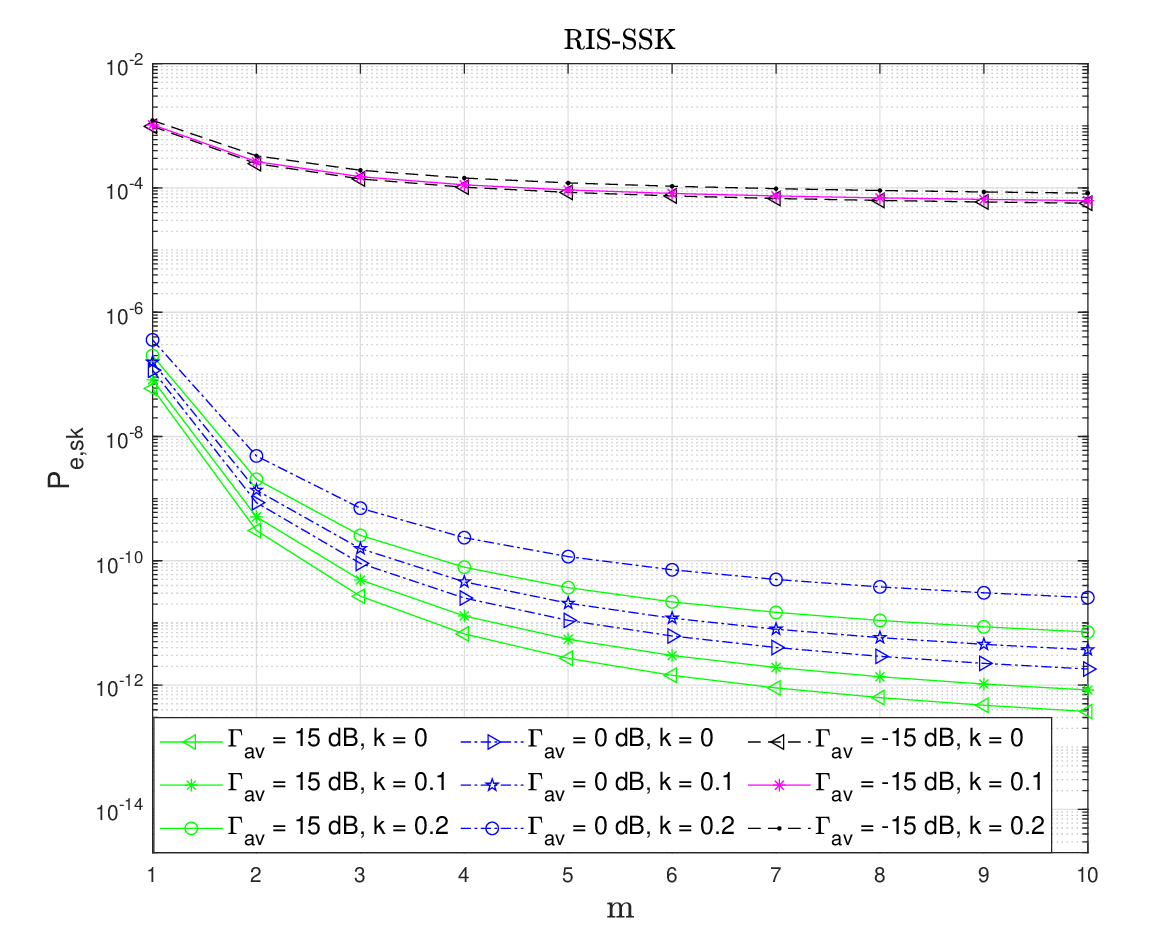}
    \caption{Variation of the PED of the SSK modulation with $m$ for different hardware impairment levels $k$. Here, $N_R = 4$ and $N = 32$.}
    \label{fig6}
\end{figure}

Fig. \ref{fig6} presents the plots of the PED versus the Nakagami shape factor $m$. In general, the performance of the communication system improves with an increase in the value of the Nakagami shape parameter due to a decrease in thee randomness of the channel or, conversely, a more deterministic nature of the channel, but to varying degrees according to the other parameters. We note the minimal impact of an increase in the shape factor on performance in scenarios with low $\Gamma_{av}$, whereas the opposite trend is observed for high $\Gamma_{av}$ levels. Conversely, we find that the performance of the communication system remains constant with changing values of $m$ at low $\Gamma_{av}$ levels. In contrast, no such saturation occurs for scenarios with higher $\Gamma_{av}$. Furthermore, the effect of varying impairment level $k$ is also more evident at higher $\Gamma_{av}$ levels, especially with changing the shape factor $m$.

\begin{figure}
    \centering
    \includegraphics[height=7.4cm, width=9.5cm]{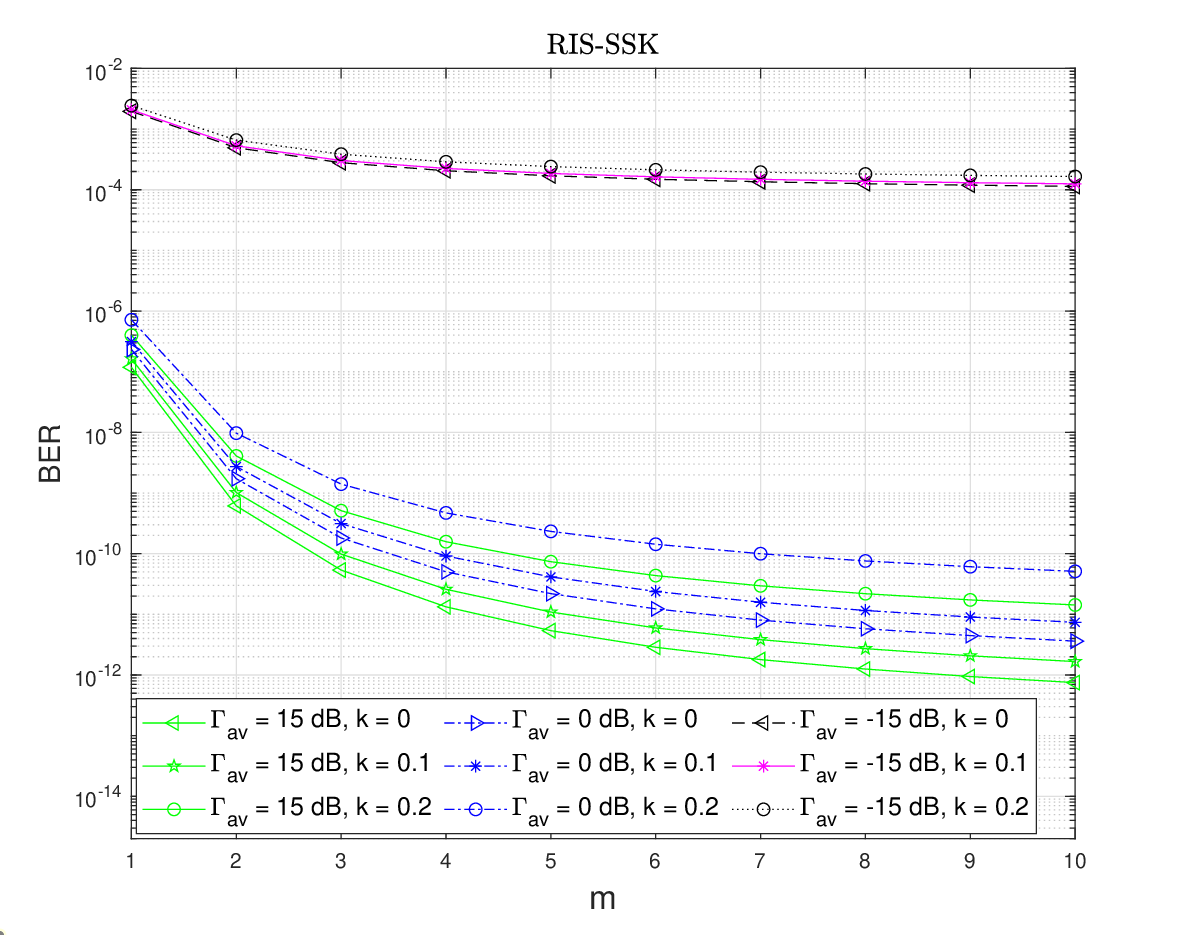}
    \caption{Variation of BER of the SSK modulation scheme with $m$ for varying values of the hardware impairment level $k$; $N_R=4$ and $N = 32$.}
    \label{fig7}
\end{figure}

Similarly, Fig. \ref{fig7} illustrates the variation of the BER with changing Nakagami shape factor $m$ for different values of the impairment level $k$, which is derived from the metric PED based on eq. (\ref{eq55}). The performance variation with changing shape factor $m$ primarily impacts systems operating at higher $\Gamma_{av}$ levels. Conversely, at lower $\Gamma_{av}$, we observe a negligible difference in performance across different impairment levels $k$ that our communication system may encounter.

\begin{figure*}
    \centering
    \includegraphics[height=6.5cm, width=18.5cm]{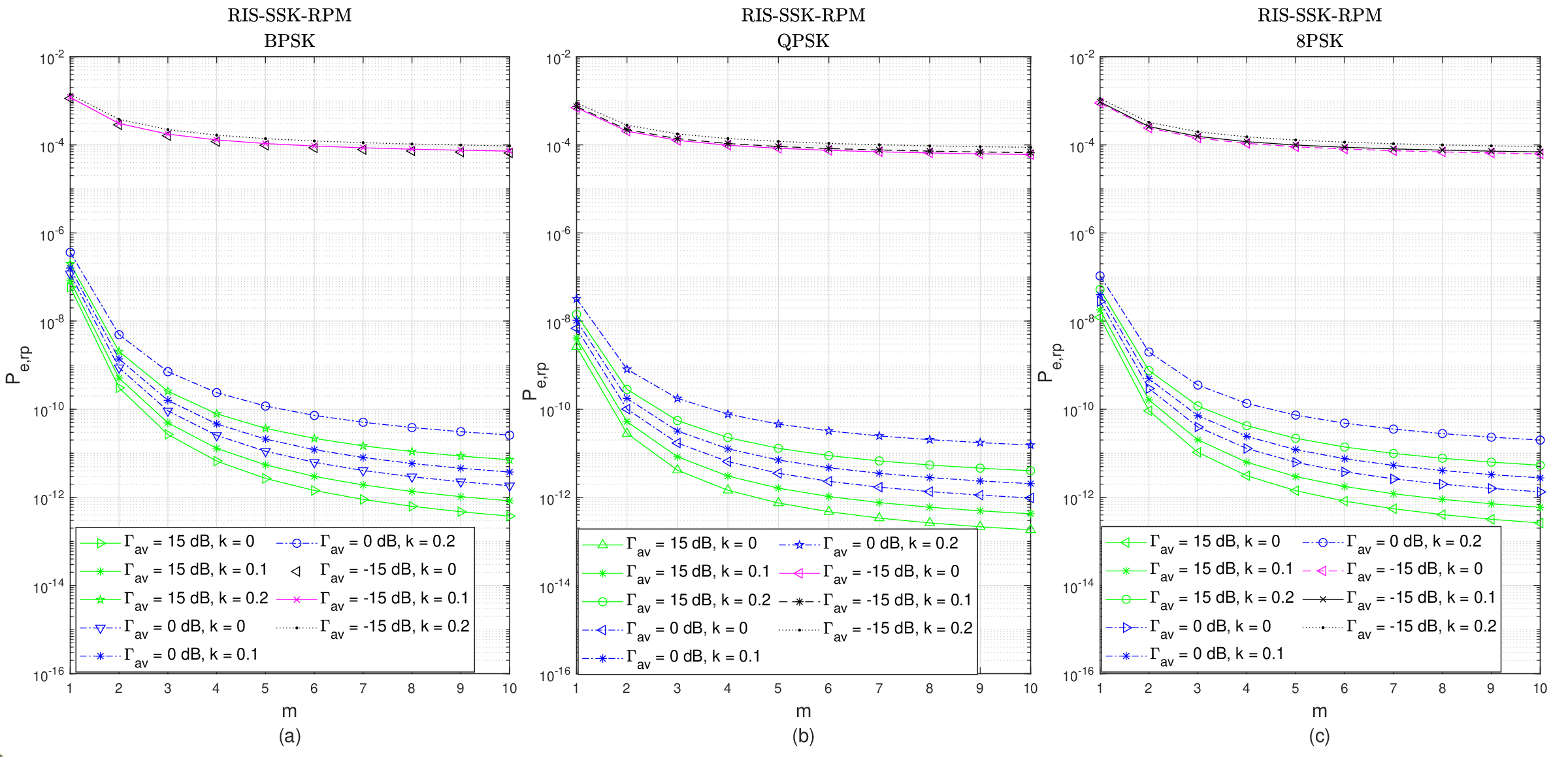}
    \caption{Variation of PED with $m$ for varying values of the hardware impairment level $k$ when $N_R = 4$ and $N = 32$, for optimized (a) SSK-RPM (BPSK), (b) SSK-RPM (QPSK), and (c) SSK-RPM (8PSK).}
    \label{fig8}
\end{figure*}

The variation of the PED with the Nakagami shape factor $m$ for RIS-assisted SSK-RPM systems employing  BPSK, QPSK, and 8-PSK modulation schemes at the RIS is presented in Fig. \ref{fig8}. In general, with an increase in the parameter $m$, we observe a saturating effect irrespective of the values of $N, N_R$, and $k$, but to a varying extent depending on parameter $\Gamma_{av}$ due to a decrease in the randomness of the channel associated with shape factor. Additionally, for constant values of $N_R$, $N$, and $k$, the performance deteriorates with an increase in the number of constellation points $M$, highlighting the constellation dependence of the PED metric. In other words, with an increase in the constellation size, the distance between the neighboring points reduces, leading to an increased PED. Furthermore, utilizing the equations (\ref{eq49}) and (\ref{eq48}), the destructive effect of the non-ideal transceivers has limited influence at lower $\Gamma_{av}$, which is evident by the early saturating effect of the PED and the converse nature of the plots observed at higher $\Gamma_{av}$ as depicted. Plots of BER versus the shape factor $m$ are not included due to a similar nature to the PED plots as observed with RIS-assisted SSK modulation.

\begin{figure}
    \centering
    \includegraphics[height=7.4cm, width=9.5cm]{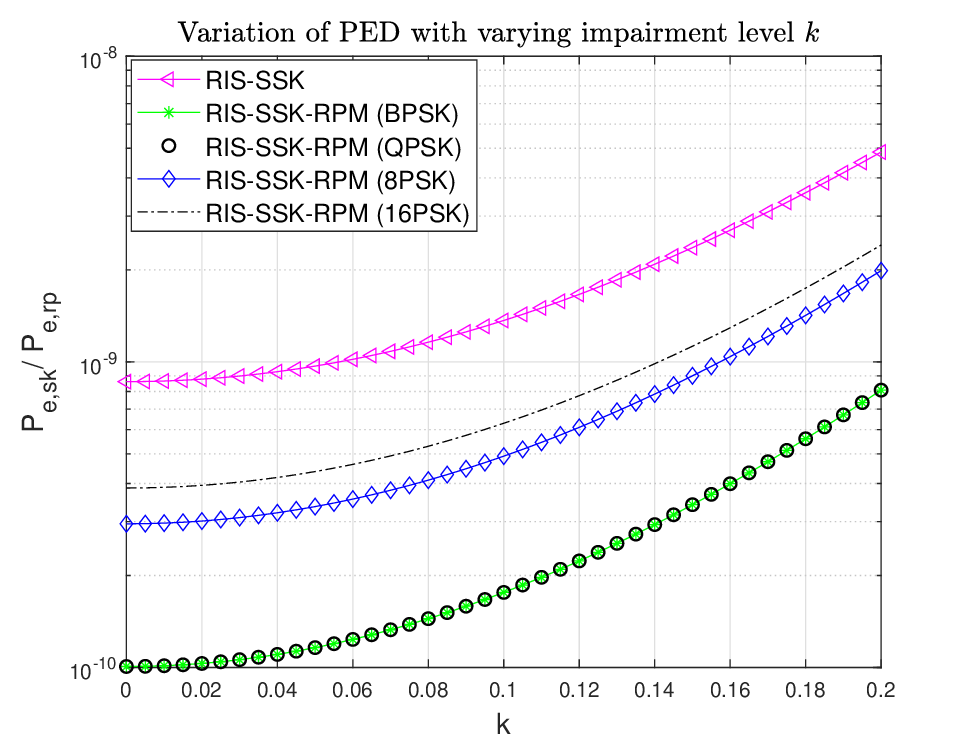}
    \caption{Variation of PED with $k$ for varying RPM constellation and SSK modulation for $N_R = 4, N = 32$, SSK modulation scheme at higher $\Gamma_{av}$ (0 dB).}
    \label{fig10}
\end{figure}

\begin{figure}
    \centering
    \includegraphics[height=7.4cm, width=9.5cm]{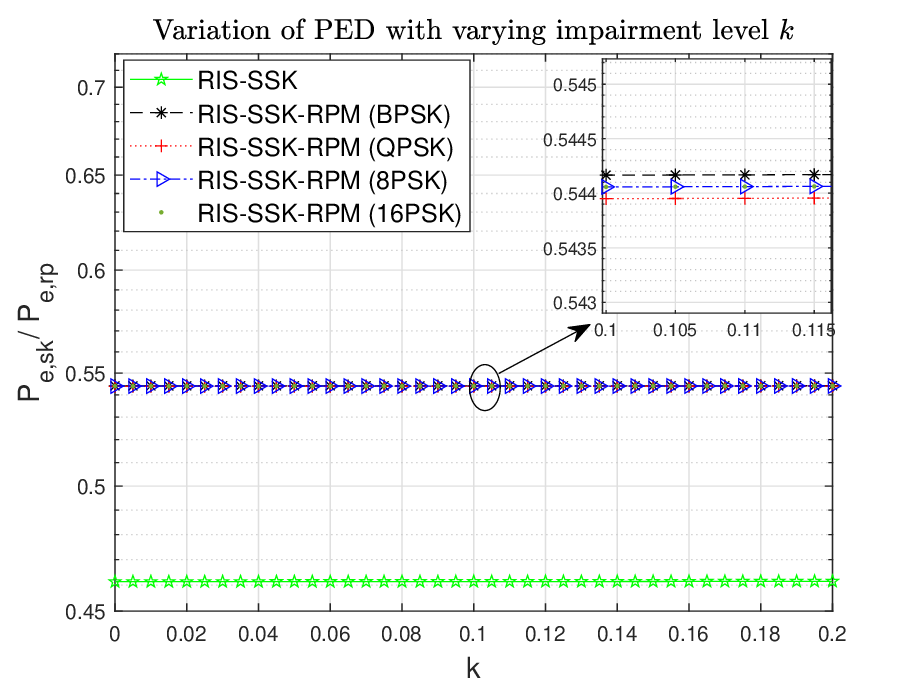}
    \caption{Variation of PED with $k$ for varying RPM constellation and SSK modulation for $N_R = 4, N = 32$, SSK modulation scheme at low $\Gamma_{av}$ (-30 dB).}
    \label{fig11}
\end{figure}

To further illustrate the performance variation of the PED metric with varying impairment level $k$ for different SSK and SSK-RPM constellations, plots at a high $\Gamma_{av}$ (Fig. \ref{fig10}) and low $\Gamma_{av}$ (Fig. \ref{fig11}) are provided. We observe from Fig. \ref{fig10} that the performance of the communication system gets affected in a non-linear manner with increasing value of the impairment level $k$. With an increase in the impairment level, the distortion noise increases, affecting the signal processing capabilities and reducing the signal quality at the receiver. Additionally, we observe that the PED is constellation-dependent and, in general, the performance degrades with an increasing number of constellation points $M$, except for QPSK and BPSK. We also observe that the performance of the RPM constellation-based communication system is superior when compared to the performance of the RIS-assisted SSK system. However, the additional complexity of implementing RPM in the RIS is a limiting factor. From Fig. \ref{fig11}, it is noted that the impairment level $k$ does not influence the performance of the RIS-assisted SSK and RIS-assisted SSK-RPM schemes, which adheres to Eqs. (\ref{eq30}) and (\ref{eq49}). Additionally, at lower values of $\Gamma_{av}$, the performance of RIS-assisted SSK is better compared to RIS-assisted SSK-RPM modulation, irrespective of the modulation used to implement RPM at the RIS. It should be noted that the limits of the impairment level $k$ are experimentally obtained in \cite{sbva213}.

\textit{Remark 5:} The numerical results above highlight the robustness of the greedy detector and the limited effect of distortion noises in terms of deviation from the ideal transceiver case. Therefore, we provide compelling arguments for the widespread application of such detectors in low-power communication scenarios such as in the IoT.
\section{Conclusion}
In this paper, we studied RIS-assisted SSK and RIS-assisted SSK-RPM schemes considering the aggregate impact of hardware impairments at the transmitters and receivers, modeled in terms of complex Gaussian distributions. The channels were modeled with i.i.d. Nakagami-$m$ distributions, and the receiver utilized a greedy detector to ascertain the index of the received diversity branch. The probability of erroneous detection (PED) of the target indexed receive diversity antenna was derived for the two schemes by using a characteristic-function-based approach, resulting in closed-form expressions, the accuracy of which was also backed by Monte-Carlo simulations. Asymptotic analysis of this reliability metric revealed that the effect of hardware impairments is limited to cases with high values of the average SNR, while little effect is observed for low average SNRs. Thus, a slight shift in the inflection point of the curves with varying impairment levels was observed. Similar trends were observed for the variation of the PED with the Nakagami shape factor, as well as for the BER analysis of the considered modulation schemes. The incorporation of path-loss and channel correlation is planned for future research to gather further insights and facilitate future deployments of RIS-assisted SSK communication systems.
\section*{Appendix A: Proof of Equation (\ref{titanic})}
By utilizing the statistics of the magnitude of the complex-valued channel gains $\alpha_{u,w}$ and $\alpha_{\hat{u},w}$, which follow i.i.d. Nakagami-$m$ distributions, and applying the central limit theorem (CLT) for scenarios involving a high number of RIS reflecting elements $N$, the statistics of the left-hand side (LHS) and right-hand side (RHS) of Eq. (\ref{qw3as}) can be defined as follows
\bsub
\beq
\left(\sqrt{E_s}+q\right) \sum_{u=1}^N \alpha_{\hat{u},w}
e^{\jmath \left(\theta_{u,w} - \theta_{\hat u,w} \right)}
+ n_{\hat u} \sim {\mathcal{CN}} \left(0, a \right) ,
\label{qw5a}
\eeq
\beq
\Re \left\{\left(\sqrt{E_s}+q\right) \sum_{u=1}^N \alpha_{u,w} + n_{u} \right\}
\sim {\mathcal{N}} \left(\mu_{1}, b_{sk} + c_{sk} \right),
\label{qw5b}
\eeq
\beq
\Im \left\{\left(\sqrt {E_s} +q\right)\sum_{u=1}^{N} \alpha_{u,w} + n_{u} \right\}
\sim {\mathcal{N}} \left(0, c_{sk} \right) ,
\label{qw5c}
\eeq
\esub
where the variables $a, b_{sk}, c_{sk}$ and $\mu_1$ are given by
\bsub
\beq
\mu_{1} = \frac{N\Gamma\left({m+\frac{1}{2}}\right)}{\Gamma\left({m}\right)}\left(\frac{E_s\Omega}{m}\right)^{\frac{1}{2}},
\eeq
\beq
a = N E_{s} \Omega + Nk^2E_{s}\Omega  + N_0,
\eeq
\beq
b_{sk}\ = NE_s\Omega\left(\!1-\!\frac{1}{m}\left(\frac{\Gamma(m+\frac{1}{2})}{\Gamma(m)}\right)^2\right),
\eeq
\beq
c_{sk} = \frac{Nk^2E_{s}\Omega}{2}\left(\!1+\!\frac{1}{m}\left(\frac{\Gamma(m+\frac{1}{2})}{\Gamma(m)}\right)^2\right) +\frac {N_{0}}{2}.
\label{eq18}
\eeq
\esub

Utilizing the statistics obtained above, we can now compute the PPED as
\beq
\Pr \left\{{ X_{sk} < Y_{sk} }\right \} = \int\limits_{-\infty }^{\infty } \Big (1 - F_{Y_{sk}}(x) \Big) f_{X_{sk}}(x) \text {d} x,
\label{ab1}
\eeq
where $F_y\left(\cdot\right)$ and $f_X\left(\cdot\right)$ are the cumulative distributive function (CDF) and the PDF, respectively. The RVs $X_{sk}$ and $Y_{sk}$ are utilized as alternate representations for $|z_w|^2$ and $|z_{\hat{w}}|^2$, respectively. The RV $X_{sk}$ can be modeled as a combination of Gaussian RVs as
\beq
X_{sk} = \left|{ \left ({W_{0} + W_{1} }\right) + \jmath W_{2} }\right |^{2} = \left ({W_{0} + W_{1} }\right)^{2} + W_{2}^{2},
\eeq
where $W_0,W_1$ and $W_2$ are independent Gaussian RVs such that
$W_0 \sim {\mathcal{N}} \left(\mu_{1}, b_{sk} \right)$ and $W_1,W_2 \sim {\mathcal{N}} \left(0,c_{sk} \right)$. Using the statistics obtained in Eq. (\ref{eq18}), the characteristic function of the RV $X_{sk}$  can be expressed as
\beqarr
{\Psi _{X_{sk}}} (\jmath \omega)
\!\!\!\! &=& \!\!\!\!
{\mathbf {E}} \left [{ e^{\jmath \omega X_{sk}} }\right] \nn \\
&=& \!\!\!\!
\frac {\exp \left \{{ \frac {\jmath \omega \mu_{1} ^{2}}{1-2\jmath \omega (b_{sk}+c_{sk})}}\right \}} {\big (1-2\jmath \omega (b_{sk}+c_{sk})\big)^{\frac {1}{2}}\big (1-2\jmath \omega c_{sk}\big)^{\frac {1}{2}}}.
\label{eq53}
\eeqarr
Additionally, the CDF of the RV $Y_{sk}$ can be obtained using the statistics obtained in Eq. (\ref{eq18}) as
\beq
F_{Y_{sk}}\left(x\right) = 1 - \exp\left\{- \frac{x}{a}\right\}.
\label{eq23}
\eeq
Therefore, substituting (\ref{eq23}) into (\ref{ab1}), followed by simplification, we obtain
\beqarr
\Pr \left \{{ X_{sk} < Y_{sk} }\right \}
\! \! \! \! &=& \! \! \! \!
\int\limits_{0}^{\infty } \exp \left \{{- \frac {x}{a} }\right \} f_{X_{sk}}(x) \text {d} x \, \nn \\
&=& \! \! \! \! \undb{E} \left[ e^{\jmath \omega X_{sk}} \right]\big|_{\jmath \omega = -\frac{1}{a}} \, .
\label{eq55a}
\eeqarr
Subsequently, substituting (\ref{eq53}) into (\ref{eq55a}), and performing algebraic simplifications, we reach the closed-form expression for the PPED as given in Eq. (\ref{titanic}), which completes the proof.
\section*{Appendix B: Proof of Equation (\ref{eq23ty})}
Using the binomial series expansion, we can write
\beq
\left( 1- x \right)^{L}
= \sum_{r=0}^{L} \left(-1 \right)^{r}
\binom{L}{r} x^r \, .
\label{eq46}
\eeq
Now, integrating both sides of Eq. (\ref{eq46}), we obtain
\beqarr
&& \! \! \! \! \! \! \! \! \! \! \! \!
\int_{0}^1 \left( 1- x \right)^{L} \text{d}x
= \sum_{r=0}^{L} \left(-1 \right)^{r}
\binom{L}{r} \int_{0}^1 x^r \text{d} x \nn \\
&& \! \! \! \! \! \! \! \! \! \! \! \! \implies \left.
- \frac{\left( 1-x\right)^{L+1}}{L+1} \right|_{0}^1
= \sum_{r=0}^{L}
\frac{\left(-1 \right)^{r}}{r+1}
\binom{L}{r} x^{r+1}\big|_{0}^1 \nn \\
&& \! \! \! \! \! \! \! \! \! \! \! \! \implies \frac{1}{L+1}
= 1 - \sum_{r=1}^{L}
\frac{\left(-1 \right)^{r-1}}{r+1} \binom{L}{r} \nn \\
&& \! \! \! \! \! \! \! \! \! \! \! \! \implies \sum_{r=1}^{L}
\frac{\left(-1 \right)^{r-1}}{r+1} \binom{L}{r}
= \frac{L}{L+1} \, .
\label{eq47}
\eeqarr
This completes the proof of eq. (\ref{eq23ty}).
\section*{Appendix C: Proof of Equation (\ref{purisf})}
Since the statistics of the magnitude of the complex-valued channel gain follow i.i.d. Nakagami-$m$ distributions, we can obtain the conditional statistics of the LHS and the RHS of Eq. (\ref{eq33}) using the CLT, i.e., assuming $N \gg 1)$, as
\bsub
\beq
\left(\sqrt{E_s}+q\right) \sum_{u=1}^{N} \alpha_{\hat{u},w}
e^{ \jmath  \left(\theta_{u,w}-\theta_{\hat{u},w}\right)}
e^{\jmath \psi_n }
+ n_{\hat{u}}
\sim {\mathcal{CN}}\left(0, a \right),
\label{eq34}
\eeq
\beqarr
&& \hspace{-1.5cm}
\Re \left\{\left(\sqrt{E_s}+q\right)\left ({\sum \limits _{u=1}^{N} \alpha_{u,w}  }\right)e^{\jmath \psi _{n}}  + n_{u} \right\}
\nn \\
&& \hspace{3cm}
\sim {\mathcal{N}} \left(\mu_{h1}, b_{rp} + c_{rp} \right),
\label{eq35}
\eeqarr
\beqarr
&& \hspace{-1.5cm}
\Im \left\{\left(\sqrt{E_s}+q\right)\left ({\sum \limits _{u=1}^{N} \alpha_{u,w}  }\right)e^{\jmath \psi _{n}}  + n_{u} \right\}
\nn \\
&& \hspace{3cm}
\sim {\mathcal{N}} \left(\mu_{h2}, d_{rp} + c_{rp} \right),
\label{eq36}
\eeqarr
\esub
where the variables $\mu_{h1}$, $\mu_{h2}$, $b_{rp}$, $c_{rp}$, and $d_{rp}$ are given by
\bsub
\beq
\mu_{h1} = N\sqrt{\frac{E_s\Omega}{m}}
\frac{\Gamma \left(m+\frac{1}{2} \right)}
{\Gamma \left( m \right)}  \sin \psi_n ,
\eeq
\beq
\mu_{h2} = N\sqrt{\frac{E_s\Omega}{m}}
\frac{\Gamma \left(m+\frac{1}{2} \right)}
{\Gamma \left( m \right)}  \cos \psi_n ,
\eeq
\beq
b_{rp} = N E_s\Omega \left(1- \frac{1}{m}
\left( \frac{\Gamma \left(m+\frac{1}{2} \right)}
{\Gamma \left( m \right)} \right)^2 \right) \sin^2 \psi_n,
\eeq
\beq
c_{rp} = \frac{Nk^2E_s\Omega}{2} \left(1+ \frac{1}{m}
\left( \frac{\Gamma \left(m+\frac{1}{2} \right)}
{\Gamma \left( m \right)} \right)^2 \right) + \frac{N_0}{2} ,
\eeq
\beq
d_{rp} = N E_s\Omega \left(1- \frac{1}{m}
\left( \frac{\Gamma \left(m+\frac{1}{2} \right)}
{\Gamma \left( m \right)} \right)^2 \right) \cos^2 \psi_n .
\label{eq37}
\eeq
\esub
It is noted that the variable $a$ has already been defined in Eq. (\ref{eq18}). Restating the terms $X_{rp}=\left| z_{w,rp} \right|^2$ and $Y_{rp}=\left| z_{\hat{w},rp} \right|^2$, the PPED expression in Eq. (\ref{eq33}), conditioned on a given RPM angle $\psi_n$, can be obtained as
\beq
\Pr \left\{X_{rp} < Y_{rp}| \psi_n\right\}
= \int\limits_{-\infty }^{\infty }\!\!\!
\left(1 - F_{Y_{rp}} \left(x \right) \right) f_{X_{rp}}(x) \text{d} x ,
\label{eq38}
\eeq
Following a similar characteristic-function-based approach as in the case of RIS-assisted SSK modulation case, we can utilize the statistics obtained in Eqs. (\ref{eq34}), (\ref{eq35}) and (\ref{eq36}) to find the CDF and PDF of the received signal energies at the target and non-target receive diversity branches, respectively. The CDF of $Y_{rp}$ mirrors that of $Y_{sk}$ given in Eq. (\ref{eq23}). Furthermore, the RV $X_{rp}$ can be equivalently expressed in terms of a Gaussian RV as
\beq
X_{rp} =\left| W_{0}^{\zeta} + \jmath W_{1}^{\zeta}\right|^2 \, ,
\label{eq39}
\eeq
where $W_{0}^{\zeta}$ and $ W_{1}^{\zeta}$ are given by
\beqarr
W_{0}^{\zeta} = {\mathcal{N}} \left(\mu_{hX1}, b_{rp} + c_{rp} \right), \nn \\
W_{1}^{\zeta} = {\mathcal{N}} \left(\mu_{hX2}, d_{rp} + c_{rp} \right) .
\label{eq50a}
\eeqarr
Thus, the characteristic function of the RV $X_{rp}$ can be formulated using Eqs. (\ref{eq39}) and (\ref{eq50a}), as
\beqarr
&& \! \! \! \! \! \! \! \! \! \! \! \! \! \! \! \!
\! \! \! \! \! \!
{\Psi _{X_{rp}}} (\jmath \omega)
= \undb{E} \left[ e^{\jmath \omega X_{rp}} \right] \nn \\
&& \! \! \! \! \! \! \! \!
=  \frac{\exp \left\{ \frac{\jmath \omega \mu_{hX1}^{2}}
{1-2\jmath \omega \left( b_{rp} + c_{rp}\right)}+\frac{\jmath \omega \mu_{hX2}^{2}}
{1-2\jmath \omega \left( d_{rp} + c_{rp} \right)}\right\}}
{\left(1-2\jmath \omega \left( b_{rp} + c_{rp} \right) \right)^{\frac{1}{2}}
\left(1-2\jmath \omega \left( d_{rp} + c_{rp} \right) \right)^{\frac{1}{2}}}.
\label{eq41}
\eeqarr
Furthermore, utilizing the CDF of $Y_{rp}$ obtained in Eq. (\ref{eq23}), we can simplify the expression in (\ref{eq38}) as
\beqarr
 \Pr \left\{X_{rp} < Y_{rp}|\psi_n \right\}
 \! \! \! \! &=& \! \! \! \!
\int\limits_{-\infty}^{\infty} \exp \left\{-\frac{x}{a} \right\}
f_{X_{rp}}(x) \text{d} x \nn \\
&=& \! \! \! \! \undb{E} \left[ e^{\jmath \omega X_{rp}} \right]\big|_{\jmath \omega = -\frac{1}{a}} \, .
\label{eq42}
\eeqarr

Finally, substituting Eq. (\ref{eq41}) into Eq. (\ref{eq42}), and subsequently performing algebraic simplifications, we obtain the closed-form expression for the PPED conditioned on the RPM angle $\psi_n$ as shown in Eq. (\ref{purisf}). The final unconditioned probability can be obtained by unconditioning over all the possible RPM constellation points as
\beq
\Pr \left\{ \! X_{rp} < Y_{rp} \right\} = \frac{1}{M}\sum_{n=1}^{M}\Pr \left\{ \left. \! X_{rp} < Y_{rp} \right| \psi_n \right\} ,
\eeq
where $M$ is the total number of constellation points that we can utilize to transmit information from the RIS, which completes the proof.

\bibliographystyle{IEEEtran}
\bibliography{references}

\end{document}